\documentclass{elsart}
\usepackage{ifpdf}
\usepackage{graphicx,amssymb,lineno}
\ifpdf
\usepackage[%
  pdftitle={Tunneling and energy splitting in an asymmetric double-well potential},%
  pdfauthor={Dae-Yup Song},%
  pdfsubject={The preprint document class elsart},%
  pdfkeywords={quantum tunneling, energy splitting, double-well potential, WKB method},%
  pdfstartview=FitH,%
  bookmarks=true,%
  bookmarksopen=true,%
  breaklinks=true,%
  colorlinks=true,%
  linkcolor=blue,anchorcolor=blue,%
  citecolor=blue,filecolor=blue,%
  menucolor=blue,pagecolor=blue,%
  urlcolor=blue]{hyperref}
\else
\usepackage[%
  breaklinks=true,%
  colorlinks=true,%
  linkcolor=blue,anchorcolor=blue,%
  citecolor=blue,filecolor=blue,%
  menucolor=blue,pagecolor=blue,%
  urlcolor=blue]{hyperref}
\fi

\makeatletter
\def\elsartstyle{%
    \def\normalsize{\@setfontsize\normalsize\@xiipt{14.5}}
    \def\small{\@setfontsize\small\@xipt{13.6}}
    \let\footnotesize=\small
    \def\large{\@setfontsize\large\@xivpt{18}}
    \def\Large{\@setfontsize\Large\@xviipt{22}}
    \skip\@mpfootins = 18\p@ \@plus 2\p@
    \normalsize
}
\@ifundefined{square}{}{}
\makeatother

\begin{document}

\begin{frontmatter}
\title{Tunneling and  energy splitting\\ in an asymmetric double-well potential}

\author{Dae-Yup Song}
\address{Department of Physics Education,
Sunchon National University, Jeonnam 540-742, Korea}

\ead{dsong@sunchon.ac.kr}

\begin{abstract}
An asymmetric double-well potential
is considered, assuming that the minima of the wells are quadratic with a
frequency $\omega$ and the difference of the minima is close to a
multiple of $\hbar \omega$. A WKB wave function is constructed on
both sides of the local maximum between the wells, by matching the
WKB function to the exact wave functions near the classical turning points.
The continuities of the wave function and its first derivative at the
local maximum then give the energy-level splitting formula, which
not only reproduces the instanton result for a symmetric potential,
but also elucidates the appearance of resonances of tunneling
in the asymmetric potential.
\end{abstract}

\begin{keyword}
quantum tunneling, energy splitting, double-well potential
\PACS 03.65.Sq, 03.65.Ge, 82.20.Xr
\end{keyword}
\end{frontmatter}

\section{Introduction}
\label{intro}
The quantum tunneling in a double-well potential appears in a
variety of physical cases. Well-known examples
include inversion of ammonia molecule \cite{DU}, for which the
double-well is symmetric, and proton tunneling in hydrogen bonds,
for which the two wells could be unsymmetrical \cite{Unsym}. For a
symmetric potential, instanton method developed in \cite{Langer},
has been elaborated and applied to calculate energy splitting \cite{Coleman}.
The tunneling splitting calculated from the instanton method exactly agrees
with the WKB result when the quadratic connection formula is adopted,
and it has been confirmed that the result is
very accurate for large separation between the two wells
\cite{BB,ZCS,Russell}.


Quantum tunneling in asymmetric double-well potentials has also long
been considered \cite{NGBCS,RCV}, and one of the intriguing
properties in these cases is the appearance of resonances. For example,
the wave function appropriate for the false vacuum of the potential
\begin{equation}
V_D(x)=\left\{\begin{array}{ll}
          \frac{m\omega^2}{2}[(x+\alpha)^2+(\beta^2-\alpha^2)] &
              ~{\rm for}~ x< 0\\
          \frac{m\omega^2}{2}(x-\beta)^2     & ~ {\rm for}~ x\geq 0,
         \end{array}\right. \label{V_D}
\end{equation}
($\beta>\alpha>0$) can significantly tunnel to the side of $x\geq 0$
only when $\beta$ is tuned to satisfy the condition
\begin{equation}
\frac{m\omega^2}{2}(\beta^2-\alpha^2)\approx n\hbar\omega, ~~~~n
=0,1,2,3\cdots.
\end{equation}
Through various numerical calculations, it has been known that the appearance
of these resonances is not limited to $V_D(x)$, but is a general property of
tunneling in asymmetric potentials \cite{NGBCS}. In this paper, in order to
elucidate the analytic structures of the resonances, we construct the WKB
wave functions for a class of asymmetric double-well potentials.

\begin{figure}
\begin{center}
\includegraphics{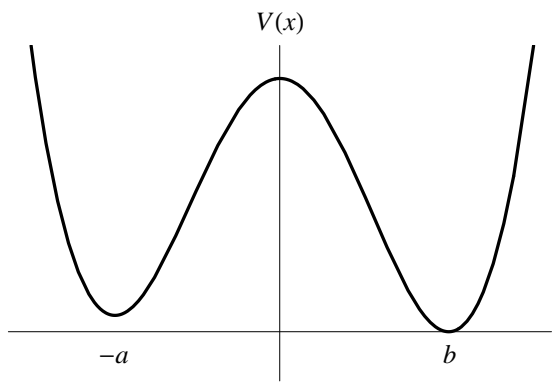}
\end{center}
\caption{ An asymmetric double-well potential $V(x)$:
$V(b)=0$, $V(-a)=(n+\epsilon)\hbar\omega$, ($n=0,1,2,\cdots$).
We assume that, for a given energy $E$, $V(x)$ is quadratic in the regions
of classical motions with the frequency $\omega$, and concentrate on the case
of $\epsilon \ll 1$.
}
\label{fig}
\end{figure}

Specifically, we consider a smooth double-well potential $V(x)$,
assuming that $V(x)$ has minima at $x=b$ and at $x=-a$ ~($a,b>0$),
and a local maximum at $x=0$. The minima are taken to be quadratic
with a frequency $\omega$, $V(b)=0$, and $V(-a)=(n+\epsilon)\hbar\omega$
(See Fig. \ref{fig}). As in the instanton
method, we are interested in the large separation between the two wells,
and consider the ground and low lying excited states of energy eigenvalue $E$
satisfying $V(0)\gg E>n\hbar\omega$. We also assume that the potential is
still quadratic near the classical turning points between the wells.

Around the minima, exact solutions to the Schr\"{o}dinger equation
are described by the parabolic cylinder functions \cite{WW}. On
both sides of $x=0$, a WKB wave function is constructed by
matching the WKB function to the asymptotic forms of the exact
solutions near the classical turning points. The continuities of
the wave function and its first derivative at the local maximum then
give the energy-level splitting formula. Though our method of requiring
continuities is very different from the instanton or WKB method in \cite{Coleman},
the splitting formula reduces to the known one in the symmetric case
\cite{Coleman,Garg,Dekker}.

In the symmetric potential, for a given energy, an approximate solution to
the Schr\"{o}dinger equation localized in left(right) well implies, by the inversion symmetry, another solution localized in the right(left) well, and this fact has
been conveniently used to evaluate energy splittings \cite{LL,Garg}.
In this paper, we also show that tunneling in the asymmetric
potential of $\epsilon=0$ can be explored by assuming the degenerate approximate solutions
to the Schr\"{o}dinger equation $\psi_R(x)$ and $\psi_L(x)$ which are localized in the right
and left wells, respectively.
By explicitly constructing $\psi_R(x)$ and $\psi_L(x)$, the splitting formula is
found from these wave functions.
Indeed it turns out that the WKB wave functions satisfying the continuities
could be written as linear combinations of $\psi_R(x)$ and
$\psi_L(x)$, while the linear combinations lift the degeneracy to make the splitting.
Therefore, a linear combination of the time-dependent WKB wave functions
gives a system which shuttles back and forth between $\psi_R(x)$ and
$\psi_L(x)$, to clearly elucidate the resonance structure of the tunneling
in the asymmetric potential of $\epsilon=0$.

This paper is organized as follows. In Section \ref{sec:continuity},
we construct the WKB wave function on both sides of the local maximum and,
by requiring the continuities at the maximum, we evaluate the energy splitting.
In Section \ref{sec:tunneling}, we find
the appropriate localized wave functions $\psi_R(x)$ and $\psi_L(x)$,
to re-obtain the energy splitting formula. We also establish a
time-dependent WKB wave function of a system shuttling back and forth.
In Section \ref{sec:concluding}, we give some concluding remarks.
Finally in Appendix A we give exact solutions for the system of $V_D(x)$
in the limit of large separation between the two wells.

\section{WKB method with continuity requirements}\label{sec:continuity}
As the results could be easily modified to incorporate
the small non-zero $\epsilon$, we start with $\epsilon=0$.
\subsection{ WKB wave function for $x\geq 0$}
 For the wave function $\psi_I(x)$ of the energy eigenvalue
$E=(\nu+n+\frac{1}{2})\hbar\omega$
around the right minimum, the Schr\"{o}dinger equation is written as
\begin{equation}
-\frac{\hbar^2}{2m}\frac{d^2}{dx^2}\psi_I
+\frac{m\omega^2}{2}(x-b)^2\psi_I=\hbar\omega(\nu+n+\frac{1}{2})\psi_I.
\end{equation}
By introducing
\begin{equation}
l_{ho}=\sqrt{\frac{\hbar}{m\omega}},
\end{equation}
and $z_R={\sqrt{2}(x-b)}/{l_{ho}}$, we rewrite the equation as
\[\frac{d^2\psi_I}{dz_R^2}+(\nu+n+\frac{1}{2}-\frac{z_R^2}{4})\psi_I=
0,\]
to obtain
\begin{equation}
\psi_I(x)=C_R D_{\nu+n}(z_R)=C_R
D_{\nu+n}(\frac{\sqrt{2}(x-b)}{l_{ho}}), \label{I pcf}
\end{equation}
where $C_R$ is a constant and $D_{\nu+n}$ denotes the parabolic
cylinder function \cite{WW}. Bearing in mind that we wish to construct
a normalizable wave function, we choose the solution in (\ref{I pcf}) so that
$\int_b^\infty |\psi_I(x)|^2 dx$ is finite if the
expression of $\psi_I(x)$ is valid for $x>b$. Asymptotic expansions
of the parabolic cylinder function are well-known. For large and
negative $z$ ($z\ll -1$and $z\ll -|k|$), we have
\begin{eqnarray}
D_k(z) &\sim&
e^{-\frac{z^2}{4}}z^k\left[1-\frac{k(k-1)}{2z^2}+\cdots\right]\cr
&&-\frac{\sqrt{2\pi}}{\Gamma(-k)}e^{k\pi i}e^{\frac{z^2}{4}}z^{-k-1}
\left[1+\frac{(k+1)(k+2)}{2z^2}+\cdots\right].~~~~~ \label{pcf
asymp}
\end{eqnarray}

For $x>0$, a classical turning point may be written as
\begin{equation}
x=b_\nu = b- \sqrt{2\nu+2n+1}~l_{ho}.
\end{equation}
In the classically forbidden region of $b_\nu-x\gg l_{ho}$, within the WKB
approximation, a solution to the Schr\"{o}dinger equation  is
\begin{equation}
\psi_{II}(x)=\frac{A_R}{\sqrt{p(x)}}e^{\int_0^x \frac{p(y)}{\hbar}
dy}+ \frac{B_R}{\sqrt{p(x)}}e^{-\int_0^x \frac{p(y)}{\hbar} dy},
\end{equation}
where $p(x)$ is defined as
\begin{equation}
p(x)=\sqrt{2m\left[V(x)-E \right]},
\end{equation}
and $A_R,~B_R$ are constants. In the region of quadratic potential
satisfying $b-x\gg b-b_\nu$, by introducing
\begin{equation}
\Phi_R(x)
=-\int_x^{b_\nu}\frac{p(y)}{\hbar} dy
=-\int_x^{b_\nu}
\frac{\left[(b-y)^2-(b-b_\nu)^2\right]^{1/2}}{l_{ho}^2}dy,
\end{equation}
we have \cite{Furry,Garg}
\begin{eqnarray}
\Phi_R(x)&=& -\frac{(b-x)^2}{2l_{ho}^2}+\frac{1}{4}(2\nu+2n+1)\cr
&&+\frac{1}{2}(2\nu+2n+1)\ln\left(\frac{2(b-x)}{b-b_\nu}\right)
+O(\left(\frac{b-b_v}{b-x}\right)^2), \label{eq:Phi_R}
\end{eqnarray}
and thus
\begin{eqnarray}
&&\psi_{II}(x)\approx\cr
&& \frac{A_R}{\sqrt{m\omega(b-x)}}
\left(\frac{2\sqrt{e}(b-x)}{b-b_\nu}\right)^{\nu+n+\frac{1}{2}}
\exp\left[\int_0^{b_\nu}\frac{p(y)}{\hbar}dy-\frac{(b-x)^2}{2l_{ho}^2}\right]
\cr &&+\frac{B_R}{\sqrt{m\omega(b-x)}}
\left(\frac{b-b_\nu}{2\sqrt{e}(b-x)}\right)^{\nu+n+\frac{1}{2}}
\exp\left[-\int_0^{b_\nu}\frac{p(y)}{\hbar}dy+\frac{(b-x)^2}{2l_{ho}^2}\right].~
\end{eqnarray}
As we are interested in the limit of large separation between the two wells
where the energy splitting is small, we introduce $\delta_l$ with an
integer $l$ ($=[\nu]$) as
\begin{equation}
\delta_l= \nu-l,
\end{equation}
so that
\begin{equation}|\delta_l|\ll 1. \label{small}
\end{equation}
In the region of the quadratic potential the wave function is also
described by $\psi_I(x)$. Making use of the asymptotic form in
(\ref{pcf asymp}), for $b-x\gg l_{ho}$, in the leading
orders we obtain
\begin{eqnarray}
&&\psi_I(x)\sim \cr
&&C_R\left[ e^{-\frac{(x-b)^2}{2l_{ho}^2}}
\left(\frac{\sqrt{2}(x-b)}{l_{ho}}\right)^{l+n}
+ \delta_l
e^{\frac{(x-b)^2}{2l_{ho}^2}}\frac{\sqrt{2\pi}(l+n)!}
{\left(\sqrt{2}l_{ho}^{-1}(x-b)\right)^{l+n+1}} \right]. \label{asym psiI}
\end{eqnarray}
By matching the asymptotic form of $\psi_I(x)$ onto that of $\psi_{II}(x)$
in this overlap region, we have
\begin{eqnarray}
A_R&=&(-1)^{l+n}~C_R\sqrt{\frac{\hbar(l+n)!~g_{l+n}}{2\sqrt{\pi}~l_{ho}}}
~e^{-\int_0^{b_l}\frac{p(y)}{\hbar}dy},\label{AR}\\
B_R&=&(-1)^{l+n+1}C_R\delta_l
\sqrt{\frac{2\pi^{3/2}\hbar(l+n)!}{l_{ho}~g_{l+n}}}
e^{\int_0^{b_l}\frac{p(y)}{\hbar}dy},\label{BR}
\end{eqnarray}
where
\begin{equation}
g_k=\frac{\sqrt{2\pi}}{k!}\left(k+\frac{1}{2}\right)^{k+\frac{1}{2}}
e^{-(k +1/2)}.
\end{equation}

\subsection{ WKB wave function for $x\leq 0$}
Since $V(x)=m\omega^2(x+a)^2/2 +n\hbar\omega$ near the minimum of the left well,
a classical turning point is given as
\begin{equation}
x=-a_\nu =-a+\sqrt{2\nu+1}~l_{ho}.
\end{equation}
In the classically forbidden region of $a_\nu +x \gg l_{ho}$, a WKB solution may be
written as
\begin{equation}
\psi_{III}(x)=\frac{A_L}{\sqrt{p(x)}}e^{-\int_x^0 \frac{p(y)}{\hbar}
dy}+ \frac{B_L}{\sqrt{p(x)}}e^{\int_x^0 \frac{p(y)}{\hbar} dy},
\end{equation}
where $A_L,~B_L$ are constants. In the region of the quadratic potential satisfying
$a+x\gg a-a_v$, from the fact that
\begin{equation}
\Phi_L(x)=-\int_{-a_\nu}^x \frac{p(y)}{\hbar} dy =-\int_{-a_\nu}^x\frac{[(a+y)^2-(a-a_\nu)^2]^{1/2}}{l_{ho}^2}dy,
\end{equation}
we obtain
\begin{eqnarray}
&&\psi_{III}(x) \approx\cr &&\frac{A_L}{\sqrt{m\omega(a+x)}}
\left(\frac{a-a_\nu}{2\sqrt{e}(a+x)}\right)^{l+\frac{1}{2}}
 \exp \left[-\int_{-a_\nu}^0 \frac{p(y)}{\hbar}dy+\frac{(a+x)^2}{2l_{ho}^2}\right]\cr
&&+\frac{B_L}{\sqrt{m\omega(a+x)}}
\left(\frac{2\sqrt{e}(a+x)}{a-a_\nu}\right)^{l+\frac{1}{2}}
\exp\left[\int_{-a_\nu}^0
\frac{p(y)}{\hbar}dy-\frac{(a+x)^2}{2l_{ho}^2} \right].~
\end{eqnarray}
With $z_L={\sqrt{2}(x+a)}/{l_{ho}}$, the Schr\"{o}dinger
equation around the left minimum is written as
\[\frac{d^2\psi_{IV}}{dz_L^2}+(\nu+\frac{1}{2}-\frac{z_L^2}{4})\psi_{IV}=
0,\] and thus
\begin{equation}
\psi_{IV}(x)=C_L D_{\nu}(-z_L)=C_L
D_{\nu}(-\frac{\sqrt{2}(x+a)}{l_{ho}}),\label{IV pcf}
\end{equation}
with a constant $C_L$. The solution in (\ref{IV pcf}) is
chosen, so that $\int_{-\infty}^{-a} |\psi_{IV}(x)|^2 dx$ is finite
if the expression of $\psi_{IV}(x)$ is valid for $x<-a$.  Around the
left minimum of $a+x\gg l_{ho}$, from (\ref{small}) and the asymptotic form in
(\ref{pcf asymp}), in the leading orders we have
\begin{eqnarray}
&&\psi_{IV}(x)\sim\cr
&&(-1)^l C_L\left[ e^{-\frac{(x+a)^2}{2l_{ho}^2}}
\left(\frac{\sqrt{2}(x+a)}{l_{ho}}\right)^l- \delta_l
~e^{\frac{(x+a)^2}{2l_{ho}^2}}\frac{\sqrt{2\pi}~l!}
{\left(\sqrt{2}l_{ho}^{-1}(x+a)\right)^{l+1}} \right]. \label{asym psiIV}
\end{eqnarray}
By matching $\psi_{III}(x)$ to $\psi_{IV}(x)$ in the overlap region,
we obtain
\begin{eqnarray}
A_L&=&(-1)^{l+1}~C_L~\delta_l~ \sqrt{\frac{2\pi^{3/2}\hbar~
l!}{l_{ho}g_l}}
e^{\int_{-a_l}^0 \frac{p(y)}{\hbar}dy},\label{AL}\\
B_L&=&(-1)^{l}~C_L\sqrt{\frac{\hbar l! g_l}{2\sqrt{\pi}l_{ho}}}
~e^{-\int_{-a_l}^0\frac{p(y)}{\hbar} dy}.\label{BL}
\end{eqnarray}

\subsection{Continuity and energy splitting}
For a smooth potential, a wave function and its first derivative  must be
continuous at $x=0$, which gives the relations
\begin{equation}
A_L=A_R, ~~~B_L=B_R. \label{con at 0}
\end{equation}
There are three unknowns $C_L,~C_R,~\delta_l$ in these two
equations, while another equation may come from the normalization of
the wave function. From $A_L/B_L=A_R/B_R$, making use of
Eqs.~(\ref{AR},\ref{BR},\ref{AL},\ref{BL}), we obtain
\begin{equation}
\delta_l^2=\frac{g_l~g_{l+n}}{(2\pi)^2}~
\exp\left[-2\int_{-a_l}^{b_l}\frac{p(y)}{\hbar} dy\right],\label{delta}
\end{equation}
which indicates that the splitting of the energy level $\Delta_l$
is given by
\begin{equation}
\Delta_l=\sqrt{g_l~
g_{l+n}}\frac{\hbar\omega}{\pi}\exp\left[-\int_{-a_l}^{b_l}
\frac{p(y)}{\hbar}dy \right]. \label{Delta_l}
\end{equation}
For a symmetric potential of $n=0$, (\ref{Delta_l}) exactly
agrees with the result in \cite{Garg,Dekker}.

Expression in (\ref{Delta_l}) is not easy to use, because the
integrand in the exponential is close to a singularity near the
limits. By introducing
\begin{equation}
I_a=\int_{-a}^0\sqrt{2m[V(y)-n\hbar\omega]}dy,~
~~~I_b=\int_0^b\sqrt{2mV(y)}dy,   \label{integral1}
\end{equation}
and
\begin{eqnarray}
\gamma_a&=&\int_0^a
\left(\frac{\sqrt{m\omega^2}}{\sqrt{2\left[V(y-a)-n\hbar\omega\right]}}
  -\frac{1}{y} \right)dy,  \cr
\gamma_b&=&\int_0^b
\left(\frac{\sqrt{m\omega^2}}{\sqrt{2V(b-y)}}
  -\frac{1}{y} \right)dy, \label{integral2}
\end{eqnarray}
the splitting is written for a general potential as
\begin{equation}
\Delta_l=\hbar\omega\frac{\sqrt{2}e^{-(I_a+I_b)/\hbar}}{\sqrt{\pi(l+n)!~l!}}
\left(\frac{\sqrt{2}~a e^{\gamma_a}}{l_{ho}}\right)^{l+1/2}
\left(\frac{\sqrt{2}~b
e^{\gamma_b}}{l_{ho}}\right)^{l+n+1/2}.~~~\label{Delta_l2}
\end{equation}
The expression in (\ref{Delta_l2}) can be conveniently used to
find that our formula reduces to the known one in \cite{Dekker}
for a symmetric potential.

\subsection{For a non-zero $\epsilon$}
The above formalism can be modified to include
non-zero $\epsilon$, as far as $\delta_l-\epsilon \ll 1$. In this
case, without a change in $\psi_I(x)$ and $\psi_{II}(x)$, the modifications
of $\psi_{III}(x)$ and $\psi_{IV}(x)$ are obtained by replacing $\delta_l$
with $\delta_l-\epsilon$ (or, $\nu$ with $\nu-\epsilon$). Due to the changes
in Eqs. (\ref{asym psiIV},\ref{AL}), the continuity requirements then give
the relation
\begin{equation}
\delta_l (\delta_l -\epsilon)
=\left( \frac{\Delta_l}{2\hbar \omega}\right)^2. \label{delta epsilon}
\end{equation}
If $\Delta_l^\epsilon$ denotes the energy splitting in the presence
of $\epsilon$, (\ref{delta epsilon}) implies
\begin{equation}
\Delta_l^\epsilon=\sqrt{\Delta_l^2+(\hbar\omega\epsilon)^2}. \label{Delta-epsilon}
\end{equation}

\section{An alternative method with localized wave functions}\label{sec:tunneling}
For $\epsilon=0$, if we assume two states of the normalized real wave
functions $\psi_R(x)$ and $\psi_L(x)$ with energy $E_0$ as mentioned in
Section \ref{intro}, the Hamiltonian in this two-state subspace may be
given by
\begin{equation}
H=\left(\begin{array}{cc}
        E_0&\Delta/2\\
        \Delta/2&E_0 \end{array}\right), \label{mHamiltonian}
\end{equation}
where the small tunneling splitting $\Delta$ is written as
\begin{equation}
\Delta=2\left|\int_{-\infty}^\infty \psi_L(x)
\left(-\frac{\hbar^2}{2m}\frac{d^2}{dx^2}+V(x)\right)\psi_R(x)dx\right|.
\end{equation}
The two eigenstates of the Hamiltonian
are given by $(\psi_R(x)\pm\psi_L(x))/\sqrt{2}$. From the Schr\"{o}dinger equations
for these eigenstates, and the equations
\[ \left[-\frac{\hbar^2}{2m}\frac{d^2}{dx^2}+V(x)\right]\psi_i(x)=E_0\psi_i(x)~~
~(i=R,L),\] and from the requirements
\begin{equation}
\int_0^\infty (\psi_R(x))^2dx\approx 1,~~~\int_{-\infty}^0
(\psi_L(x))^2dx\approx 1,~~~\int_{0}^\infty
\psi_L(x)\psi_R(x)dx\approx 0,
\end{equation}
we find
\begin{equation}
\Delta\approx \frac{\hbar^2}{m}\left|\psi_L(0)\psi_R'(0)
-\psi_R(0)\psi_L'(0)\right|,
\end{equation}
with $\psi_i'(x)={d\psi_i(x)}/{dx}$ $~(i=R,L)$, which is a generalization of the method
used for the symmetric case \cite{Garg,LL}.

In the classically forbidden region,
we may write
\begin{equation}
\psi_R(x)=\frac{N_R}{\sqrt{p(x)}}
e^{\int_0^x\frac{p(y)}{\hbar}dy},~~
\psi_L(x)=\frac{N_L}{\sqrt{p(x)}} e^{-\int_0^x\frac{p(y)}{\hbar}dy},
\label{waves on wells}
\end{equation}
where $N_R$ and $N_L$ are constants. From these expressions of
$\psi_R(x)$ and $\psi_L(x)$, by assuming that the validity condition
for the WKB approximation
\begin{equation}
\left|\frac{d}{dx}\frac{\hbar}{p}\right|\ll 1
\end{equation}
is satisfied at $x=0$, we obtain
\begin{equation}
\Delta\approx 2\frac{\hbar}{m}\mid N_LN_R\mid.\label{TS}
\end{equation}

For $E_0=(l+n+\frac{1}{2})\hbar\omega$, near the right-hand well, $\psi_R(x)$ would be
accurately described by the $(l+n)$th harmonic oscillator
eigenfunction. This description holds well into the forbidden
region, and, for $(b-x)/ l_{ho}\gg 1$, we may write
\begin{equation}
\psi_R(x)\approx
\frac{\exp\left[-\frac{(x-b)^2}{2l_{ho}^2}\right]}{\sqrt{\sqrt{\pi}~l_{ho}(l+n)!}}
\left(\frac{\sqrt{2}(x-b)}{l_{ho}}\right)^{l+n} . \label{eq:RHO}
\end{equation}
On the other hand, making use of (\ref{eq:Phi_R}), we can find the asymptotic
expansion form of $\psi_R(x)$ of (\ref{waves on wells}) in the overlap region.
By matching the asymptotic form onto the expression in (\ref{eq:RHO}),
we obtain
\begin{equation}
N_R=(-1)^{l+n}\frac{\sqrt{\hbar ~g_{l+n}}}{\sqrt{2\pi}~l_{ho}}
\exp\left[-\int_0^{b_\nu}\frac{p(y)}{\hbar}dy\right].
\end{equation}
Similarly, by matching the asymptotic form of $\psi_L(x)$ onto that
of the $l$th excited harmonic oscillator state near the left-hand well,
we have
\begin{equation}
N_L=\frac{\sqrt{\hbar ~g_{l}}}{\sqrt{2\pi}~l_{ho}}
\exp\left[-\int_{-a_\nu}^0\frac{p(y)}{\hbar}dy\right].
\end{equation}
By plugging these explicit forms of $N_R$ and $N_L$ into
(\ref{TS}), for $E_0=(l+n+\frac{1}{2})\hbar\omega$, we confirm that $\Delta$
reduces to $\Delta_l$.

If (\ref{con at 0}) is satisfied, after some algebra,  we find that
$\psi_{II}(x)$ and $\psi_{III}(x)$ of $\epsilon=0$ can be merged, in the
classically forbidden region, into
\begin{equation}
\psi_{WKB}^{\pm}(x)=\frac{\sqrt{2\pi} A_R l_{ho}}{\sqrt{\hbar g_{l+n}}}
\exp \left[\int_0^{b_\nu}\frac{p(y)}{\hbar}dy\right]
\left[(-1)^{l+n}\psi_R(x)\mp\psi_L(x)\right], \label{WKB in both}
\end{equation}
namely, $\psi_{WKB}^{\pm}(x)$ becomes $\psi_{II}(x)$ for $x\geq 0$, and
$\psi_{III}(x)$ for $x\leq 0$, where $\psi_{WKB}^{+}(x)$ ($\psi_{WKB}^{-}(x)$) is
given when we choose $\delta_l >0$ ($\delta_l <0$) in (\ref{delta}). Thus this
alternative method is in fact equivalent to the WKB method of requiring
the continuities. We also note that $\psi_{WKB}^{+}(x)$ has a node in the
classically forbidden region between the wells, while
$\psi_{WKB}^{-}(x)$ has no node in the same region.

A (unnormalized) time-dependent WKB solution is given as
\begin{eqnarray}
\psi(x,t)&=&e^{-i\omega\left(n+l+\frac{1}{2}+|\delta_l|\right)t}\psi_{WKB}^{+}(x)
+e^{-i\omega\left(n+l+\frac{1}{2}-|\delta_l|\right)t}\psi_{WKB}^{-}(x)\cr
&=&2\frac{\sqrt{2\pi} A_R l_{ho}}{\sqrt{\hbar g_{l+n}}}
\exp \left[-i\omega(n+l+\frac{1}{2})t+\int_0^{b_\nu}\frac{p(y)}{\hbar}dy\right]\cr
&&\times\left[(-1)^{l+n}\cos\left(\frac{\Delta_l}{2\hbar}t\right)\psi_R(x)
   -i\sin\left(\frac{\Delta_l}{2\hbar}t\right)\psi_L(x)\right].
\end{eqnarray}
This last form shows clearly that the system shuttles back and forth
between $\psi_R(x)$ and $\psi_L(x)$ with the frequency $\Delta_l/\hbar$.

In order to include $\epsilon$, let us consider a slight modification
of the potential around the left well so that $V(-a)$ changes from $n\hbar\omega$
to $(n+\epsilon)\hbar\omega$. While $\psi_R (x)$ would still be an appropriate solution
of the new system with energy $E_0$, we introduce $\psi_L^\epsilon (x)$
as an approximate solution with energy $E_0+\epsilon\hbar\omega$ localized in the 
left well. If we confine our attention on
the two state subspace described by $\psi_L^\epsilon (x)$ and $\psi_R (x)$,
the Hamiltonian of the new system is analogous to that of a particle in a magnetic
field \cite{RMP}. We note that this analogy can be used to derive (\ref{Delta-epsilon})
within the approximation that $\psi_L^\epsilon (x)$ is the same with
$\psi_L (x)$. If $\psi^\epsilon(x,t)$ is a solution in this subspace with $\psi^\epsilon(x,0)=\psi_R(x)$, the maximum of the probability of $|\int_{-\infty}^\infty(\psi_L^\epsilon(x))^*\psi^\epsilon(x,t)|^2$ 
during the time-evolution is $\Delta_l^2/(\Delta_l^2+\epsilon^2)$, which indicates that 
the resonance peaks in tunneling have the Lorentzian shape.

\section{Concluding remarks}\label{sec:concluding}

Energy splitting formula has been obtained for the asymmetric
double-well potential, by assuming that the potential is quadratic near the
minima. As has been well known for the symmetric case, we expect that the
splitting formula given here would be very accurate for the large separation
between wells, which needs to be confirmed through numerical calculations.
If we could add a linear term $sx$ to the
potential $V(x)$ of $\epsilon=0$ with a controllable constant $s$,
$V(x)+sx$ has two minima at $x=b-{s}/{m\omega^2}$ and at $x=
-a-{s}/{m\omega^2}$. Since the difference of the minima is given as
$n\hbar \omega -s(a+b)$, in the light of numerical
results \cite{NGBCS}, the tunneling would be significant only if $s$
is close to a multiple of ${\hbar \omega}/{(a+b)}$. It would be of great
interest to realize the asymmetric system with controllable constants.
As a final remark, for $n=0$ and $\epsilon=0$, we note that
$\Delta_{2l}/\Delta_0=\left(2abe^{\gamma_a
+\gamma_b}/l_{ho}^2\right)^{2l}/(2l)!$ which shows the
quasi-Weierstrassian nature of the tunneling spectrum \cite{Dekker}.
In this asymmetric case, thus, the tunneling behavior of
an initially squeezed wave packet is erratic, and the trajectory of
the expected position of the wave packet has a fractal structure.

\section*{Appendix A}
For the potential $V_D(x)$ in (\ref{V_D}), WKB method may not be
applicable, since the potential is not differentiable at $x=0$.
In this case, however, exact wave functions could be
written in terms of the parabolic cylinder functions on both sides
of $x=0$. From the continuities of the wave function
and its first derivative at $x=0$, assuming
$\frac{m\omega^2}{2}(\beta^2-\alpha^2)=(n+\epsilon)\hbar\omega,$
we find that the eigenstate of an energy eigenvalue
$(\nu+n+\epsilon+\frac{1}{2})\hbar\omega$ exists
if the condition
\begin{equation}
D_\nu\left(-\frac{\sqrt{2}~\alpha}{l_{ho}}\right)
D_{\nu+n+\epsilon}'\left(-\frac{\sqrt{2}~\beta}{l_{ho}}\right)
= -D_\nu'\left(-\frac{\sqrt{2}~\alpha}{l_{ho}}\right)
D_{\nu+n+\epsilon}\left(-\frac{\sqrt{2}~\beta}{l_{ho}}\right)
\label{DD}
\end{equation}
is satisfied. (\ref{DD}) can be solved in the limit of
$\alpha,~\beta\gg l_{ho}$ and $\epsilon \ll1$. Making use of the asymptotic expansion
of (\ref{pcf asymp}), in this limit we obtain
\begin{equation}
\delta_l^2 +\left(r(R_l-L_l)+\epsilon\right)
\delta_l-R_lL_l-\epsilon r L_l=0, \label{qdelta}
\end{equation}
where
\begin{eqnarray}
&&R_l=\frac{(\sqrt{2}\beta/l_{ho})^{2(n+l)+1}}{\sqrt{2\pi}(l+n)!}
e^{-\beta^2/l_{ho}^2},~~~L_l=\frac{(\sqrt{2}\alpha/l_{ho})^{2l+1}}{\sqrt{2\pi}l!}
e^{-\alpha^2/l_{ho}^2},\cr &&r=\frac{\beta-\alpha}{\alpha+\beta}.
\end{eqnarray}

(\ref{qdelta}) implies that the energy splitting
is given as
\begin{equation}
\hbar\omega\sqrt{4R_lL_l+\epsilon^2+ 2\epsilon r (R_l+L_l)+ r^2
(R_l-L_l)^2}. \label{adouble}
\end{equation}
If we formally use the
formulas in Eqs.~(\ref{integral1},\ref{integral2}) by replacing $V(x)$ with $V_D(x)$,
$\Delta_l$ coincides with the splitting of (\ref{adouble}) in the symmetric
case ($n=0,~\epsilon=0$).

\end{document}